\documentclass[a4paper,12pt]{article}
\usepackage{cancel}
\usepackage{ulem}
\usepackage{amsfonts}
\usepackage{amssymb}
\usepackage{graphicx}
\usepackage{amsmath,bm}
\usepackage{enumerate}
\usepackage{mathtools}
\usepackage{tikz} \usetikzlibrary{calc}
\usepackage{subfig}
\usepackage{xcolor}
\usepackage{float}
\usepackage{color}
\usepackage{here}
\usepackage{cite}

\usepackage{mathrsfs}
\usepackage{float,epsfig}
\usepackage{dcolumn}
\usepackage{pgfplots}
\usepackage{graphicx}
\usepackage{bm}
\usepackage{amsmath,amssymb,amsthm}
\usepackage[colorlinks=true,linkcolor=blue,citecolor=red]{hyperref}
\newcommand{\be}{\begin{equation}}
\newcommand{\ee}{\end{equation}}
\newcommand{\bea}{\setlength\arraycolsep{2pt} \begin{eqnarray}}
\newcommand{\eea}{\end{eqnarray}}

\def\0{{\sst{(0)}}}
\def\1{{\sst{(1)}}}
\def\2{{\sst{(2)}}}
\def\3{{\sst{(3)}}}
\def\4{{\sst{(4)}}}
\def\5{{\sst{(5)}}}
\def\6{{\sst{(6)}}}
\def\7{{\sst{(7)}}}
\def\8{{\sst{(8)}}}
\def\sst#1{{\scriptscriptstyle #1}}

\definecolor{lime}{HTML}{A6CE39}
\newcommand{\orcidicon}{%
    \begin{tikzpicture}
    \draw[lime, fill=lime] (0,0)
        circle [radius=0.16]
        node[white] {{\fontfamily{qag}\selectfont \tiny ID}};
    \draw[white, fill=white] (-0.0625,0.095)
        circle [radius=0.007];
    \end{tikzpicture}   \hspace{-2mm}
}
\newcommand\orcidAdil{{\href{https://orcid.org/0000-0001-7623-5541}{\orcidicon}}}
\newcommand\orcidHajar{{\href{https://orcid.org/0000-0001-9510-4248}{\orcidicon}}}

\makeatletter \@addtoreset{equation}{section}

\usepackage{multirow}
\begin{document}

\title{\bf \Large
	Superentropic AdS Black Hole Shadows }
\author{   \small A. Belhaj\orcidAdil\!\!\footnote{a-belhaj@um5r.ac.ma}, H. Belmahi\orcidHajar\!\!\footnote{hajar\_belmahi@um5.ac.ma}, M. Benali\orcidAdil\!\!\footnote{mohamed\_benali4@um5.ac.ma}\footnote{ Authors in alphabetical order.}
	\hspace*{-8pt} \\
	{\small  D\'{e}partement de Physique, Equipe des Sciences de la mati\`ere et du rayonnement,
		ESMaR}\\ {\small   Facult\'e des Sciences, Universit\'e Mohammed V de Rabat, Rabat,  Morocco} }\maketitle

\begin{abstract}
		{\noindent}We  study  shadow aspects of superentropic black holes in four dimensions. Using Hamilton-Jacobi  formalism,  we first  get  the  null  geodesic equations.  In  the celestial  coordinate  framework relying on  fixed positions of observers,   we   investigate  the   shadow behaviors  in terms   of the mass and  the cosmological scale   variation parameters. Among others, we   obtain   ellipse shaped geometries contrary to  usual black hole solutions.
Modifying the  ordinary   relations describing geometrical observables, we discuss   the size and  the shape  deformation  parameters of  these  non-trivial geometric  forms. Due to horizonless limits associated with  certain mass values, we  explore   the shadow of  the naked singularity of such black holes.\\
\\
{\bf Keywords}: Superentropic, Cosmological constant,  Shadows, Geometrical  observables.
	\end{abstract}
\newpage
\tableofcontents

\section{Introduction}
Recently, there has been a great interest in black hole physics in connection with various gravity theories. This has been encouraged by remarkable efforts of Event Horizon Telescope collaborations. These international  activities have provided interesting imaging of supermassive black holes at the center of galaxy 87 \cite{111,222,333}. Remarkably,  various works have dealt  with thermodynamic and optical aspects of such fascinating objects. In  Anti-de Sitter (AdS) geometries,  non-trivial results associated with many transitions in black holes have been approached \cite{D22,1222}. In particular, the Hawking-Page transition has  been investigated in different backgrounds including string-theory and related models \cite{77,78}. It has been shown that the study of such a transition unveil certain universalities  reported in  \cite{79,K1}. Parallelly  to these efforts, optical behaviors have been investigated by  considering  the  deflection angle of lights  and the  shadow geometry \cite{K,G,1,O,M1,MM,Belhaj:2021tfc}. Concretely, the black hole shadow behaviors have been examined in different theories including M-theory and superstring theory in the presence of D-brane objects\cite{H,M}. The visualization of the shadow cast using the null geodesic equations  has been  completed by the study of geometrical observables. These quantities provide  information about the involved size and the shape. For non-rotating black holes, it has been revealed  that the shadow has a circular geometry where its size can be controlled by the mass and other parameters including the charge and dark energy  \cite{55}. This geometric manifestation  is distorted for ordinary rotating black holes, to exhibit  either  the D-shape or the cardioid  geometries  \cite{H,M, C1,C2}. In addition, it has been shown that  the shadow behaviors depend  also on other parameters including  the brane number  and  a cosmological scale  \cite{H,555}.\\
More recently,  several efforts have been devoted to    explore and study     the  pulsar SGR J174-2900 near  supermassive black holes  SgrA*\cite{C3}.   This provides   data on    the  horizon and the  horizonless  of events around   such  compact objects.     It has been shown that   for  rotating  black holes or a naked singularity,  the emission frequency of  such a  pulsar could be   modified\cite{C4}.
A special interest has been put on superentropic  black holes as a fascinate  solution  with non compact horizon topologies which could  exceed  entropy of maximum bounds \cite{R,G1}.    Imposing  appropriate  limits, the superentropic  black hole  is  considered as  an ultraspinning	limit of the  Kerr-Newman-AdS solution \cite{14,13}.
 Many investigations have been  conducted  by approaching such a class of black holes. More precisely, the thermodynamic aspects have been studied in  \cite{12,15}.  In particular, it has been established an interplay between  the  thermodynamics of these black holes and the usual  ones using the ultra-spinning approximation limits \cite{16,12}.   Moreover,   it has been  shown that the  superentopic   version of certain usual black holes  cannot  be derived from  ultra-spinning ones  \cite{166,WU}.\\
 The aim of this work  is to contribute to these activities by investigating shadow aspects of superentropic black holes in four dimensions. Concretely,  we first derive the associated  null  geodesic equations of motion using the  Hamilton-Jacobi  formalism. Fixing the observer positions,  we then obtain the celestial  coordinates needed to  illustrate the corresponding  shadow behaviors by varying the mass and  the cosmological scale parameters. Among others, we find  elliptic  shaped geometries contrary to  usual black holes.
Modifying the  ordinary   relations describing geometrical observables, we study  the size and  the shape  deformation  parameters of the obtained   non-trivial geometric  forms. Due to horizonless limits for certain mass values, we explore the shadow of  naked singularity of such black hole solutions.\\
The organization of this work is  follows. In section 2, we give a  concise  review on superentropic black hole solutions. In section 3, we elaborate the associated   geodesic  photon  equations using the  Hamilton-Jacobi  analysis. In section 4, we investigate the shadow  behaviors. The last section is devoted to conclusions and open questions.

\section{Superentropic  black hole models}
In this section, we give a concise review on   of superentropic  black hole models  in four dimensions, being  new solutions of the Einstien-Maxwell  equations. These solutions could be obtained from Kerr-Newman-$AdS_{4}$ black holes where the rotating parameter                                                                                                                                                                                                                                                                                                                                                                                                                                                                                                                                                                                                                                                                                                                                                                                                                                                                                                                                                                                                                                                                                                                                           is replaced by the AdS length scale    using an appropriate approximation limit\cite{13}. According to \cite{16,12},  the line element of  the associated  metric  reads as
\begin{eqnarray}
ds^{2}=&-&\frac{\Delta_{r}}{\Sigma} \left(dt - {\ell} \sin^2\theta d\phi \right)^2 +\Sigma  \left(\frac{dr^2}{\Delta_{r}}
+\frac{d\theta^2}{\sin^2\theta} \right)\\
\nonumber
&+& \frac{ \sin^4\theta}{\Sigma} \left( \ell dt -{(r^2+\ell^2)}d\phi \right)^2,
\end{eqnarray}
where $\ell$ is the AdS radius length  being connected to the cosmological constant $\Lambda$ via the relation
 \begin{equation}
 \label{er}
{\Lambda}=-\frac{3} {\ell^{2}}.
\end{equation}
The involved quantities are expressed as follows 
 \begin{equation}
\Sigma=r^2+\ell^2\cos^2\theta, \hspace{0.5 cm} \Delta_r=(\ell+\frac{r^2}{\ell})^2-2mr+q^2.
\end{equation}
It is noted that  $m$ and $q$ are the mass and the charge parameters, respectively. The local  coordinate $\phi$,   being a  noncompact direction, should be compactified   as follows $\phi \sim \phi + \alpha $,  where  $\alpha$ is a   dimensionless parameter. The later  has been  identified with a  new chemical potential $K$\cite{16}. Solving the equation $\Delta_r=0$, one can get   the large root $r_+$,   corresponding to  the existence of the black hole horizon. Such a solution   requires a  constraint on  the mass parameter      given by
\begin{equation}
\label{ES }
m\geqslant 2r_c(\frac{r_c^2}{\ell^2}+1),
\end{equation}
where one has $r_c^2=\frac{\ell^2}{3}\left[(4+\frac{3}{\ell^2}q^2)^{1/2}-1\right]$.   This constraint  being   a relationship between  $m$,  $\ell$ and $q$  black hole parameters provides two kinds of superentropic black hole solutions. Precisely, the first one  is associated with the existence of the horizon, where the constraint is verified. However,  the second one  corresponds to a naked singularity where $\Delta_r$ takes complex values. In this way,  the range of parameters are reduced up to such a constraint.   The thermodynamics of such black hole solutions have  been investigated in  many places including in \cite{12}.  In particular,  certain quantities  have been approached in terms of usual  charged and rotating black holes.   Here,  however, we attempt to unveil  the associated optical  behaviors. Concretely, we  study the shadow geometric forms casted by   ultra-spinning black holes in four dimensions.
\section{Geodesic equations of motion  }
Before presenting shadow aspects, we first elaborate the geodesic equations of motion using test particles. In particular,
the equations of photons surrounding   the black hole  horizons can be  obtained     by exploiting the   Hamilton-Jacobi  formalism. According to  \cite{A2},  one uses the following  relation
\begin{equation}
\label{ks1}
\frac{\partial S}{\partial \tau}+\frac{1}{2}g^{\mu\nu}\frac{\partial S}{\partial x^\mu}\frac{\partial S}{\partial x^\nu}=0,
\end{equation}
where $\tau$ is the affine parameter satisfying  the geodesic equations, and where $g_{\mu\nu}$ is the associated metric. $S$ denotes  the Jacobi action  which reads as
\begin{equation}
\label{ks2}
S=-Et+L\phi+S_r(r)+S_\theta(\theta),
\end{equation}
where $E$ and $L$ are the total energy and the  angular momentum of the photons, respectively. They are related   to the four-momentum $p_\mu$ via the relations $E=-p_t$ and $L=p_\phi$.  It is noted that $S_r(r)$ and  $S_\theta(\theta)$ are two  functions depending on $r$ and $\theta$ variables, respectively. Implementing the separation method and the Carter constant, the complete null geodesic equations  can be  given in terms of the impact parameters expressed  as follows
\begin{equation}
\label{xe}
 \xi=\frac{L}{E}, \hspace{1.5cm}\eta=\frac{\mathcal{K}}{E^2},
\end{equation}
where $\mathcal{K}$ is a separable constant\cite{A1}. To get the  desired  relations, certain computations should be performed. Indeed, they generate  the following null geodesic equations
\begin{align}
\Sigma \frac{d  \, t}{d \tau}& =  E \left[ \frac{A\left(r^2 +\ell^2 \right) }{\Delta_r} +  \frac{ \ell \left(\xi  - \ell \sin^2 \theta \right) }{ \sin^2\theta} \right], \\
\label{r}
\Sigma  \frac{d  \, r}{d \tau} &=\sqrt{\mathcal{R}(r}),\\
 \Sigma\frac{d  \, \theta}{d \tau} & =\sqrt{\Theta(\theta)},\\
 \label{phi}
\Sigma \frac{d  \, \phi}{d \tau} &= E \left[ \frac{A\;\ell}{\Delta_r} + \frac{ \xi-\ell \sin^2 \theta}{\sin^4 \theta }  \right],
\end{align}
where the radial $\mathcal{R}(r)$ and   the polar $\Theta(\theta)$ functions of motion are given by
\begin{eqnarray}
\mathcal{R}(r) & = & E^2\left[ A^2- \Delta_r\left({(\ell-\xi)^2}+\eta \right)  \right],\\
\Theta(\theta) & = & E^2 \left[ \eta \sin^2\theta  - \cos^2\theta\left( {\xi^2\cos^2\theta}-2\xi\right ) \right],
\end{eqnarray}
and where   one has taken $A=  \left(r^2 +\ell^2 \right)- \ell \xi$.
\section{Shadow behaviors}
In this  section, we study the shadow aspects  of superentropic   black holes. First, we consider the  existing  horizons. Then, we deal with the naked singularity behaviors associated with non-existing horizons.
\subsection{Shadows of existing horizons}
\begin{figure}[ht!]
		\begin{center}
		\centering
\includegraphics[scale=.6]{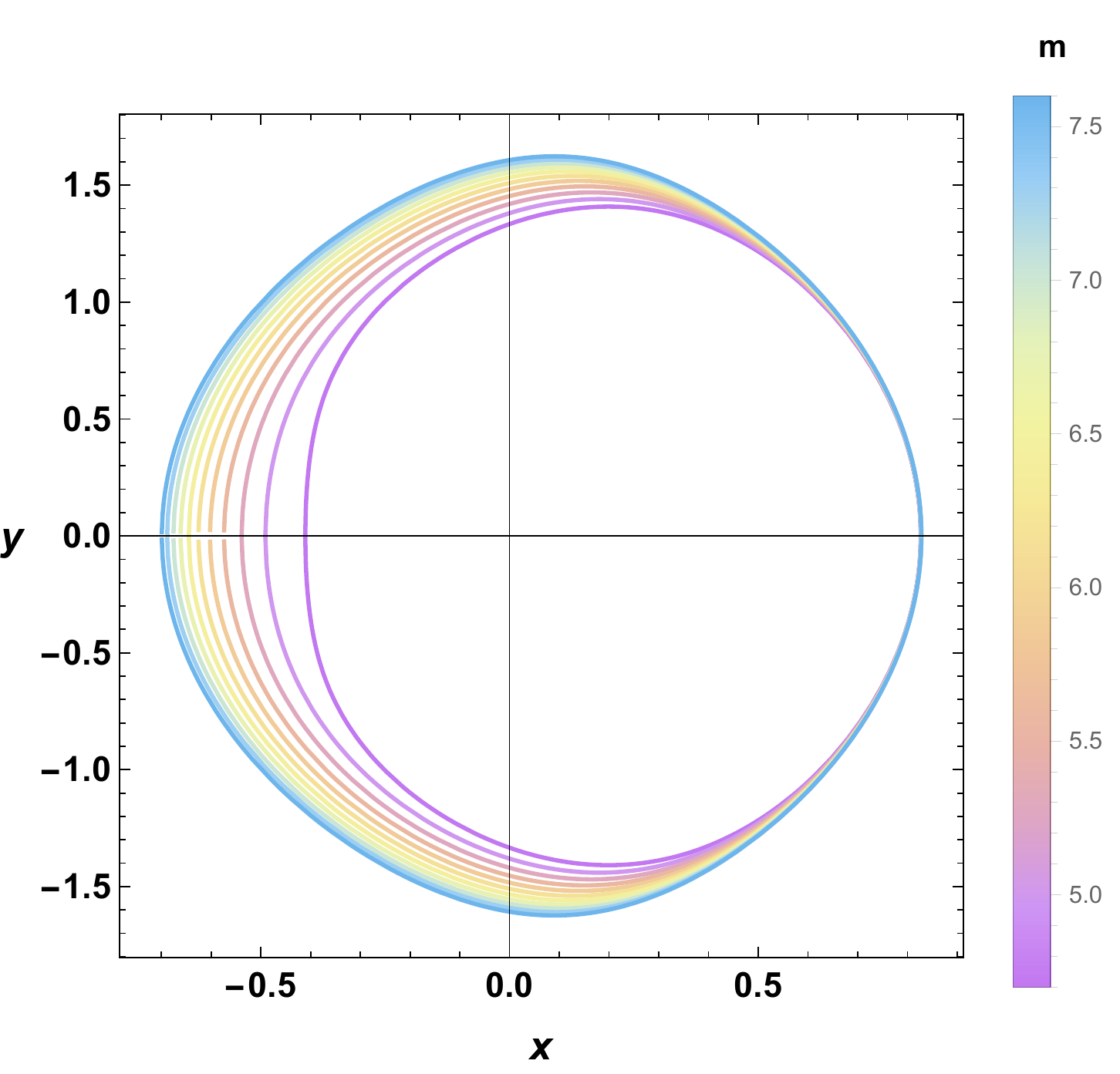}
\caption{{\it \footnotesize {Shadow behaviors  of superentropic AdS black holes for different values of  $m$ in the horizon domain  existence,  by taking $\ell=3$ and  $q=0$. The observer is positioned at $r_{ob}=100$ and $\theta_{ob}=\frac{\pi}{2}$}}}.
\label{f1}
\end{center}
\end{figure}

\begin{figure}[ht!]
		\begin{center}
		\centering
\includegraphics[scale=.6]{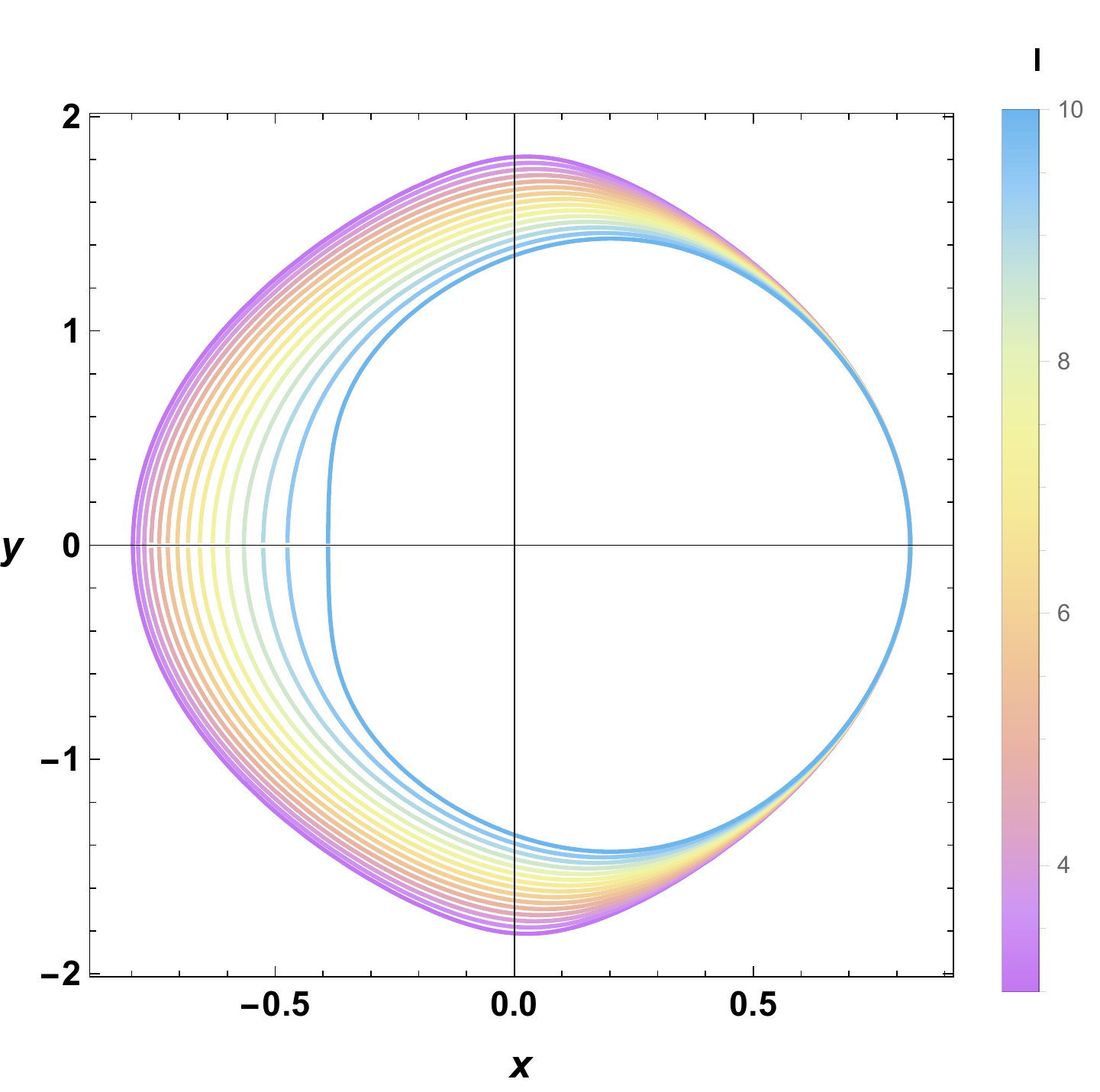}
\caption{{\it \footnotesize {Shadow behaviors  of superentropic AdS black holes for different values of  $\ell$,  by taking $m=15.4$ and  $q=0$. The observer is located  at $r_{ob}=100$ and $\theta_{ob}=\frac{\pi}{2}$.}}}
\label{f20}
\end{center}
\end{figure}

\begin{figure}[ht!]
		\begin{center}
		\centering
\includegraphics[scale=.6]{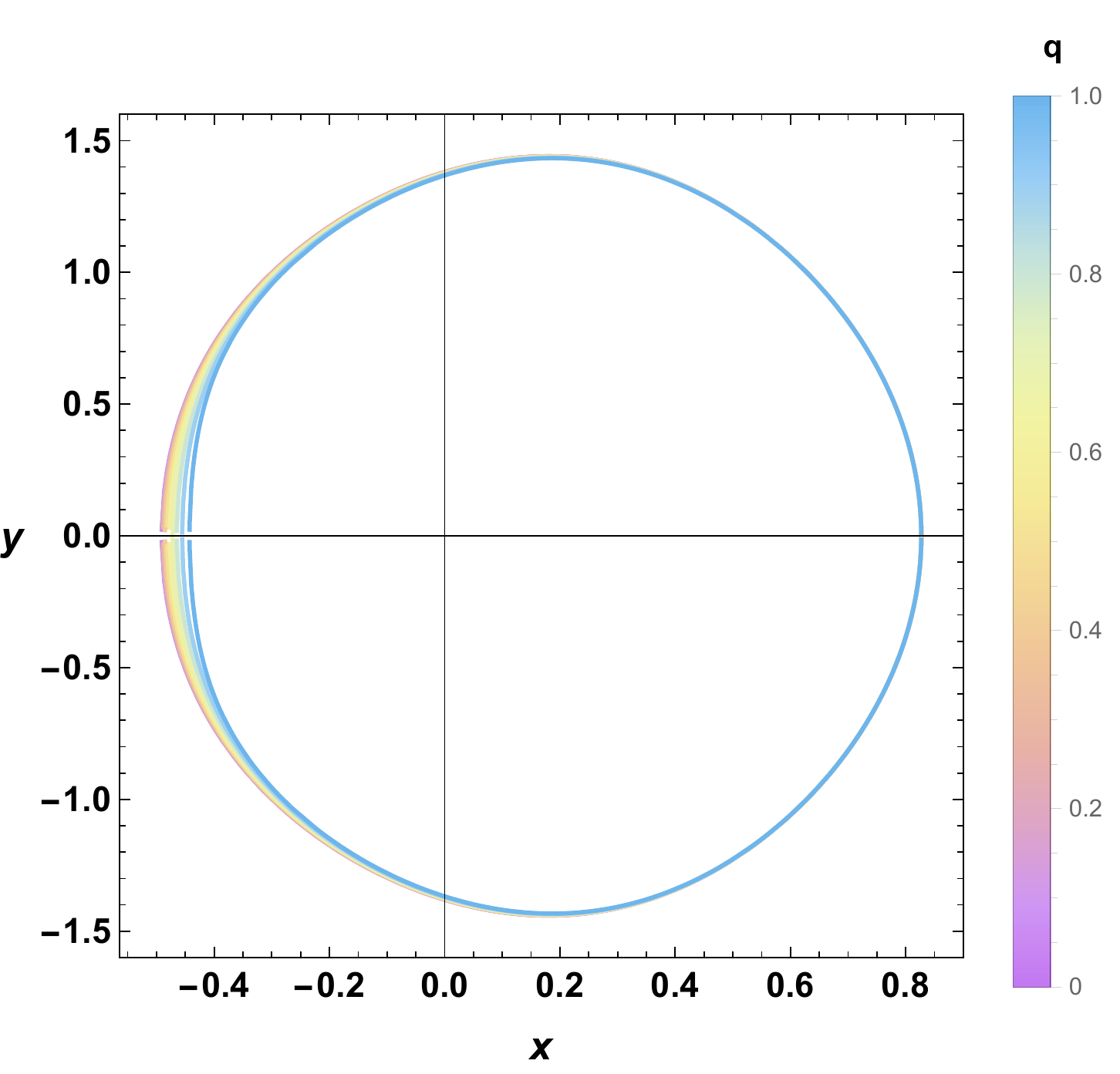}
\caption{{\it \footnotesize {Shadow behaviors  of superentropic AdS black holes for different values of  $q$,  by taking $m=5$ and  $\ell=3$. The observer is positioned at $r_{ob}=100$ and $\theta_{ob}=\frac{\pi}{2}$.}}}
\label{f21}
\end{center}
\end{figure}

For this situation, the boundary of the black hole geometric shapes can be determined from  the unstable circular orbits characterized by
\begin{equation}
\mathcal{R}(r)\Big|_{r=r_s}=\frac{d\, \mathcal{R}(r)}{d r}\Big|_{r=r_s}=0,
\end{equation}
where $r_s$ is the circular orbit radius of the photon\cite{Belhaj:2021tfc}.  Appropriate  performed calculations  give
\begin{align}
& \eta =\frac{r^2 \left(16 \ell^2 \Delta_r- \left(4 \Delta_r-r \Delta_r^{\prime}\right)^2\right)}{\ell^2{ \Delta_r^{\prime}}^2}\bigg\vert_{r=r_s},\\
& \xi=\frac{\left(r^2+ \ell^2\right) \Delta_r^\prime-4r \Delta_r  }{\ell \Delta_r^\prime} \bigg\vert_{r=r_s}.
\end{align}
It is  known,   in the presence of the cosmological constant,  that the distance $r_{ob}$ of the observer in domain of  outer communications  $(\Delta_r>0)$  should be fixed\cite{D3,D2}.  To get shadow geometries of the superentropic AdS black holes, we assume that the observer is located in the following  frame
\begin{eqnarray}
\label{e_0}
e_0 & = & \frac{ (r^2+\ell^2)\partial_t+\ell\partial_\phi}{\sqrt{\Delta_r \Sigma}} \bigg\vert_{(r_{ob}\,,\theta_{ob})},\\
\label{e_1}
e_1 & = & \frac{\sin\theta}{\sqrt{\Sigma}} \partial_\theta\bigg\vert_{(r_{ob}\,,\theta_{ob})},\\
\label{e_2}
e_2 & = &  -\frac{ \ell\sin^2{\theta}\partial_t+\partial_\phi}{\sqrt{\Sigma}\sin^2{\theta}} \bigg\vert_{(r_{ob}\,,\theta_{ob})},\\
\label{e_3}
e_3 & = &-\frac{\sqrt{\Delta_r}}{\sqrt{\Sigma}} \partial_r\bigg\vert_{(r_{ob}\,,\theta_{ob})}.
\end{eqnarray}
The timelike vector $e_0$  represents    the four-velocity of the  observer. $e_3$  indicates   the vector along the spatial direction pointing toward the center of the black hole. However,  $e_0\pm e_3$  can be considered as tangent  directions to the  one  of principal null congruences and $\theta_{ob}$ is the angle of  the observer. In particular, we consider the ray of lights  being  characterized  by the equation $
\lambda(s)=(r(s),\theta(s),\phi(s),t(s))$ and tangent to the observer  position
 \begin{equation}
\label{lambdat}
\dot{\lambda}=\dot{r}\partial_r+\dot{\theta}\partial_\theta+\dot{\phi}\partial_\phi+\dot{t}\partial_t,
\end{equation}
where the 3-vector of the spacelike   can be represented  in a basis corresponding to the  spherical coordinates.   Introducing such coordinates,   we get the tangent equation  in terms  of orthonormal tetrads and celestial coordinates as follows
\begin{equation}
\label{lambdatt}
\dot{\lambda}=\beta\left(-e_0+\sin\psi\cos\delta e_1+\sin\psi\sin\delta e_2+\cos\psi e_3\right),
\end{equation}
where $\beta$ is  a scalar factor \cite{D1,D2}. Combining  the ray of light equations and Eq.(\ref{lambdatt}),  we obtain
\begin{equation}
\label{alpha}
\beta=g(\dot{\lambda},e_0)=\frac{E}{\sqrt{\Delta_r\Sigma}}\Big(\ell\xi-(r^2+\ell^2)\Big) \bigg\vert_{(r_{ob}\,,\theta_{ob})}.
\end{equation}
An examination  reveals that the celestial coordinates $\psi$ and $\delta$  can be expressed as a function of the impact  parameters $\xi$ and $\eta$. The comparison of the coefficients $\partial_r$ and $\partial_\phi$   gives  the celestial coordinates in the terms of $\dot{r}$ and $\dot{\phi}$.  Exploiting  Eq.(\ref{r}) and Eq.(\ref{phi}), we get the celestial coordinates as a function of $\xi$ and $\eta$
\begin{eqnarray}
\label{rho}
\sin{\psi}&=&\frac{\pm\sqrt{\Delta_r\eta}}{((r^2+\ell^2)-\ell\xi)}\bigg\vert_{(r_{ob}\,,\theta_{ob})},\\
\label{psi}
\sin{\delta}&=&\frac{\sqrt{\Delta_r}}{\sin\psi}\left(\frac{(\ell-\csc^2{\theta}\xi)}{\ell\xi-(r^2+\ell^2)}\right)\bigg\vert_{(r_{ob}\,,\theta_{ob})}.
\end{eqnarray}
According to \cite{D1}, the boundary of   the shadows  can be  obtained  using  the cartesian coordinate system
\begin{eqnarray}
\label{x}
x & = & -2 \tan{(\frac{\psi}{2})}\sin{\delta},\\
\label{y}
y & = & -2 \tan{(\frac{\psi}{2})}\cos{\delta}.
\end{eqnarray}
A close inspection  shows that the shadow  behaviors depend on certain black hole parameters including the mass, the charge and the cosmological scale.  To visualize such  geometric configurations, we first  consider  neutral solutions.
In Fig.(\ref{f1}), the corresponding  shadow  aspects are plotted in such a plane by exploiting $x$ and $y$ expressions.
Concretely, we illustrate the shadow geometrical behaviors in terms of the parameter $m$  of the reduced moduli space associated with the existence of the horizon.  It follows form this figure that the shadow ellipse shaped geometry is increased by increasing the mass parameter $m$. For the values of $m$ close to a critical mass $m_c=2r_c({r_c^2}/{\ell^2}+1)$, we obtain  the D-shape  elliptic geometry. For the values of $m$  above  to  such a critical mass,  the D-shape  elliptic geometry  disappears.   This  shadow behavior  comes from the black hole  mass variation. The latter is associated with the   horizon existence as functions of   black hole parameters including $ \ell$,  $m$ and	 $q$.
It has been remarked that the geometry of the shadow is not a perfect circle arising in ordinary non-rotating black holes  \cite{F1,F10,F11,F1M,F2M}.\\ Having discussed the effect of the mass parameter on non-charged  black holes, we move now to introduce the  influence   of  the $\ell$  parameter on the shadow  geometry.  Fixing 
the  mass parameter to the value $m=15.4$ associated with $\ell=10$,   we  illustrate  the  corresponding behaviors  by varying $\ell$. An examination of  Eq.(\ref{er}) shows that  one can consider the range  $3\leqslant\ell \leqslant10$ to keep  the AdS solutions. This computation is depicted  in Fig.(\ref{f20}).
It  is observed, from this figure, that the shadow geometry size is increased by decreasing the parameter $\ell$. For the values of  the parameter $\ell$ close to the value $\ell=10$, we recover  the D-shape elliptic geometry. For  $3\leqslant\ell \leqslant9$,  however,  we obtain a generic  ellipse shaped configuration.   Fixing the mass parameter,  it  has been remarked that the  large values of  the $\ell$ parameter, associated with D-shape geometry,   can   play the same role as the rotating one   of  ordinary black hole solutions.
Now, we inspect the charge influence on such geometries by fixing  $m$ and $\ell$.  The  variation of the shadow behaviors in terms of  the charge  parameter  $q$  is represented in Fig.(\ref{f21}).    For  $0\leqslant q \leqslant1$,   it  has been remarked that  the  shadow geometry involves the same shape. This could suggest that the charge does not  involve relevant effects.
\subsection{Geometric observables}
\begin{figure}[ht!]
		\begin{center}
		\centering
\includegraphics[scale=.6]{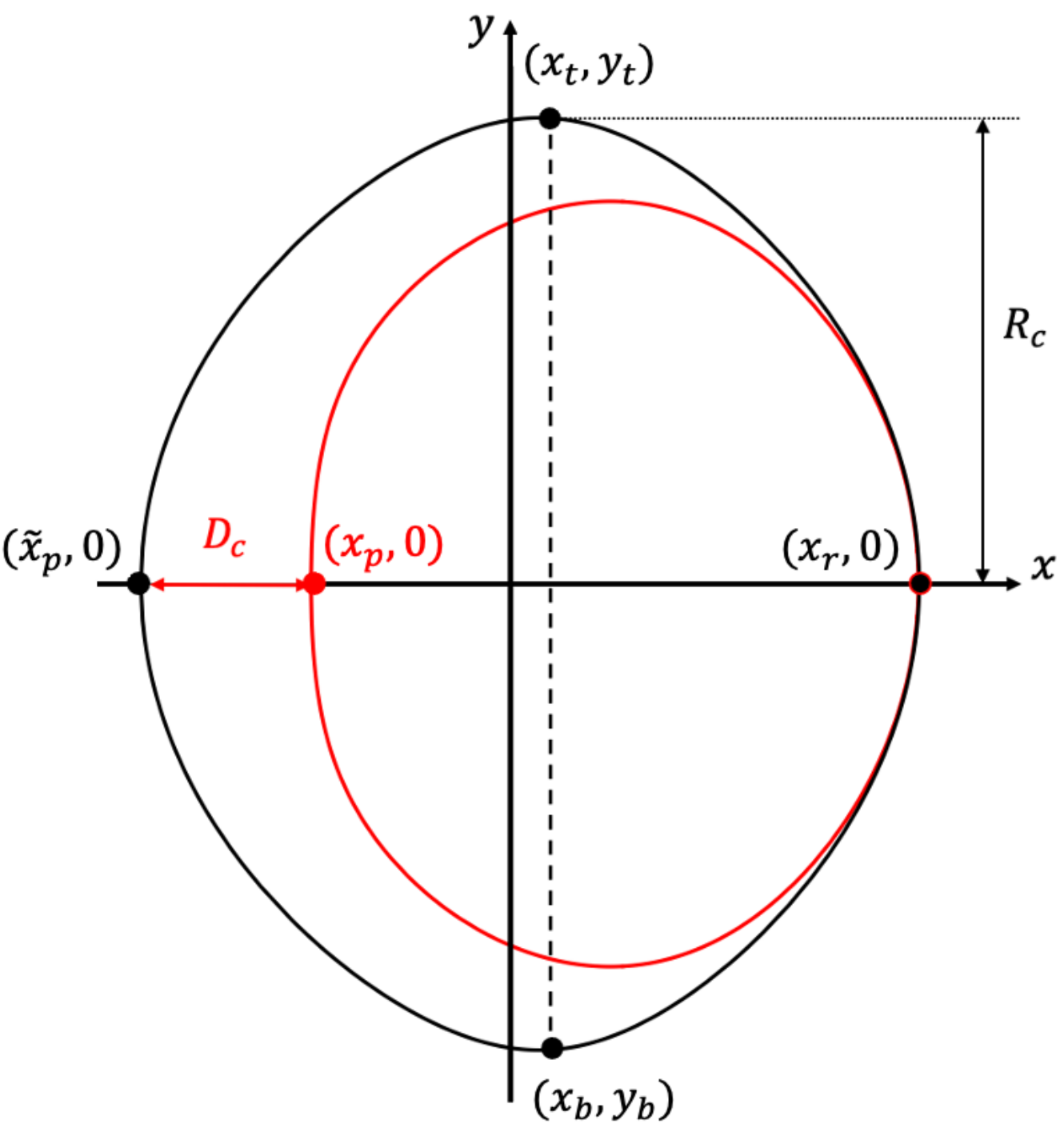}
\caption{{\it \footnotesize { Illustration of parameter deformations of shadows. }}}
\label{f221}
\end{center}
\end{figure}

\begin{figure}[ht!]
		\begin{center}
		\centering
\includegraphics[scale=.9]{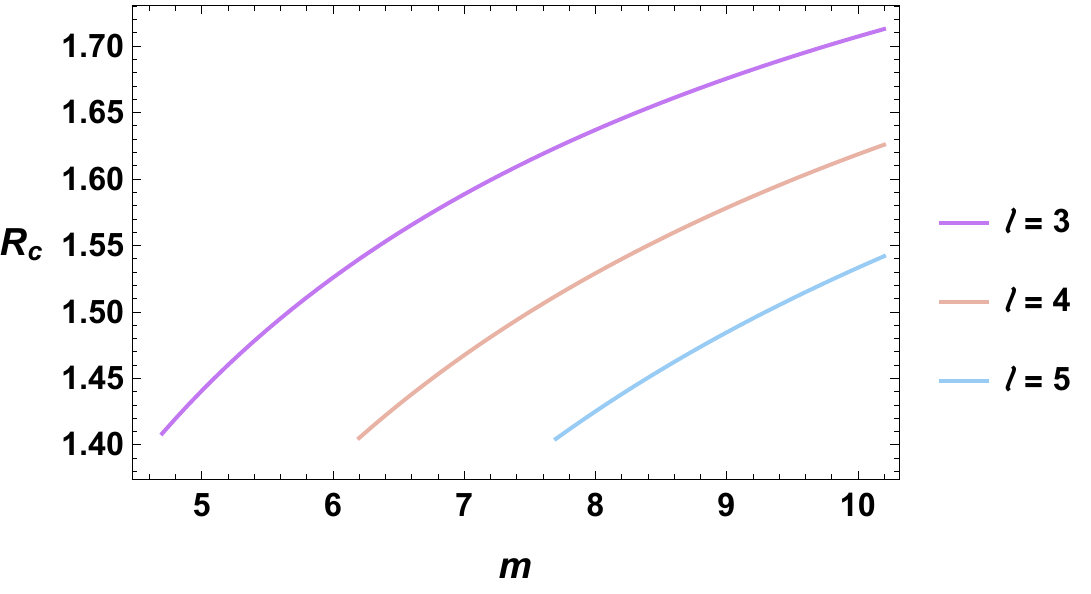}
\caption{{\it \footnotesize { Size variations of  superentropic AdS black holes for different values of  $m$, and  $\ell$  by taking $q=0$.}}}
\label{f224}
\end{center}
\end{figure}

\begin{figure}[ht!]
		\begin{center}
		\centering
\includegraphics[scale=.9]{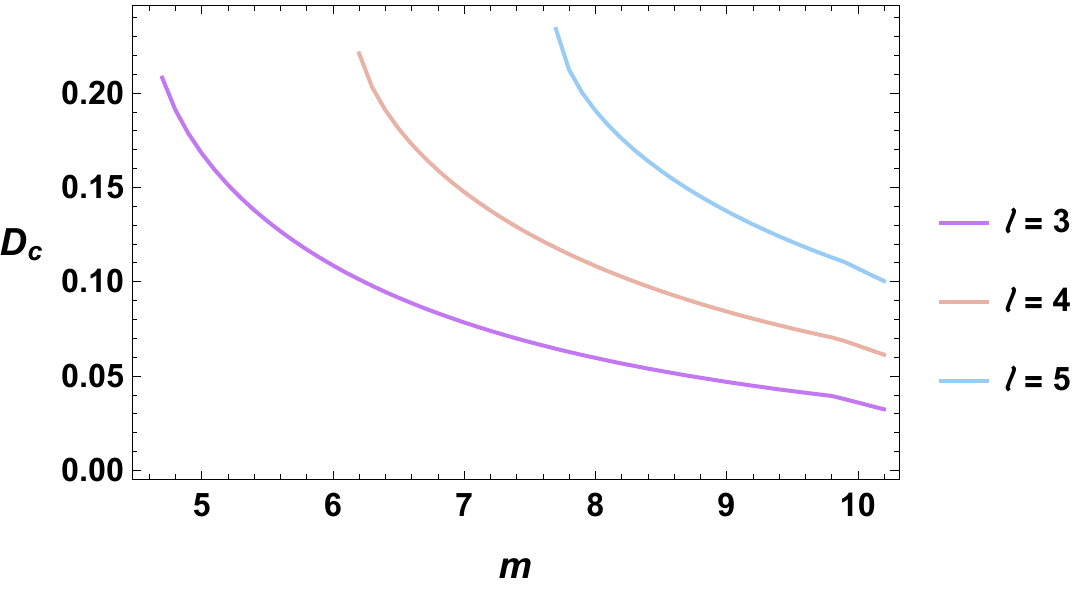}
\caption{{\it \footnotesize { Shape deformations of superentropic AdS black holes for different values of  $m$, and  $\ell$ by taking $q=0$. }}}
\label{f225}
\end{center}
\end{figure}

Here, we   discuss the geometrical observables  controlling  the size and the shape of the studied shadows.  A close examination shows  that the usual   relations could be modified  due to  elliptic geometry of shadows.    A priori, there could be many ways to do so. However, we consider here   a simple calculation.  In this way,   one has    the radius $R_c$, which represents the maximal radius of the ellipse shaped geometry,
and the distortion $D_c$  indicating the distance between the no-deformed ellipse and the D-shaped one.
   Following  \cite{hioki2009measurement,amir2016shapes},  the shadow of the black hole is characterized by three specific points.  However,
   the positions of  the  top ($x_t,y_t$) and the bottom ($x_b,y_b$) positions  have been  modified according to the present case.  The point of a
   standard  ellipse  ($\tilde{x}_p,0$)  and  the point of  the distorted shadow ellipse  ($x_p,0$) intersect  the horizontal axis   associated  with  the $x$ direction. This modification is illustrated in Fig.(\ref{f221}).
 It is worth noting that the usual relations can  be recovered by taking the circular limit. Now, we discuss such modified geometrical quantities. The associated  computations  are given in Fig.(\ref{f224}) and  Fig(\ref{f225}).
Varying the mass parameter, the $R_c$ behavior has been illustrated for various values  of $\ell$. It has been observed from Fig.(\ref{f224}) that such a parameter increases in terms of $m$. At generic values of this parameter, $R_c$ decreases with the cosmological scale $\ell$.  Fig.(\ref{f225}) shows the behavior of $D_c$ in terms of $m$. It follows from  this figure that $D_c$ decreases by increasing $m$. Contrary to $R_c$, $D_c$ increases with $\ell$ for fixed values of $m$.
\subsection{Naked singularity}
It has been shown that black holes and  naked singularities can be distinguished using different approaches including accretion disks\cite{AD}. We expect that the shadow geometric configuration can be also exploited to show such distinctions. In the vicinity of  naked singularities,  it has been suggested that  gravity effects should be dominant, which  could   be visible  from  the shadow   geometric configurations.  In this way, this new optical characteristic of  such a  singularity can also be exploited  to unveil  more data  in the future  shadow observations of the galactic center M87\cite{222}.
For such reasons, we discuss the  naked singularity shadows  of superentropic AdS black holes in four dimensions.  Inspecting the previous  shadow geometries,    the unstable spherical orbits of the photons involve  an elliptic  geometry.   In   the naked singularity, however,   such  orbits  are illustrated by  arcs.  It is  recalled  that  the naked singularity  appears  when the  largest root of  $\Delta_r=0$  takes complex values.
 Due  to the  horizon absence for certain  values of  the mass lower to   the critical mass of the superentropic AdS black holes,  the  photons which are close to both sides of the  possible  arcs can  be seen by  the observer\cite{D4,D5}.\\
 Considering  neutral solutions and taking into account the
 critical mass constraint, we vary the cosmological parameter $\ell$ with  the same range of  the  mass parameter in the  domain  of the horizonless. In this way,     one should have $m<m_c$ for generic values of  $\ell$.
 \begin{figure}[!htb]
		\begin{center}
		\centering
			\begin{tabbing}
			\centering
			\hspace{8.6cm}\=\kill
			\includegraphics[scale=.5]{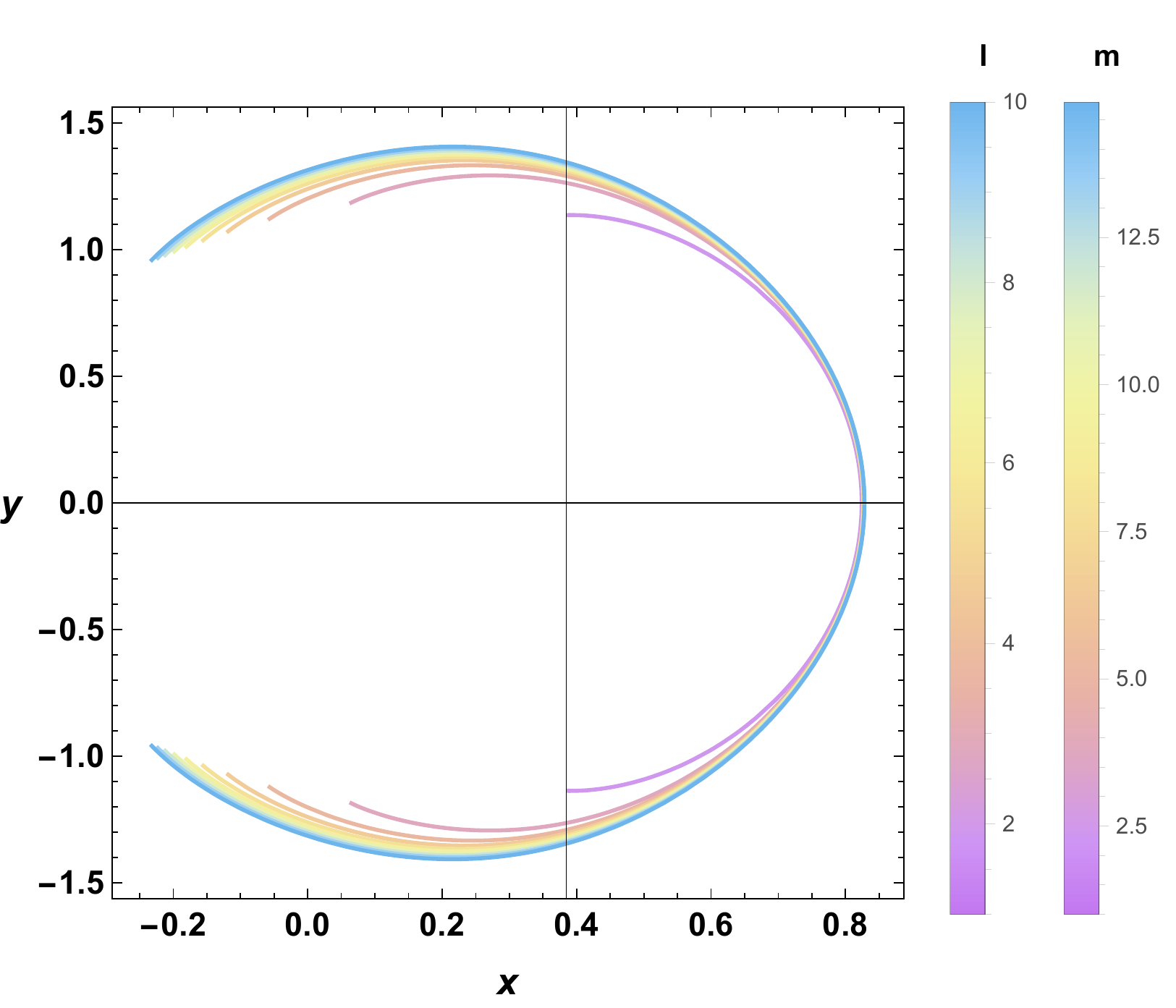} \>
			\includegraphics[scale=.68]{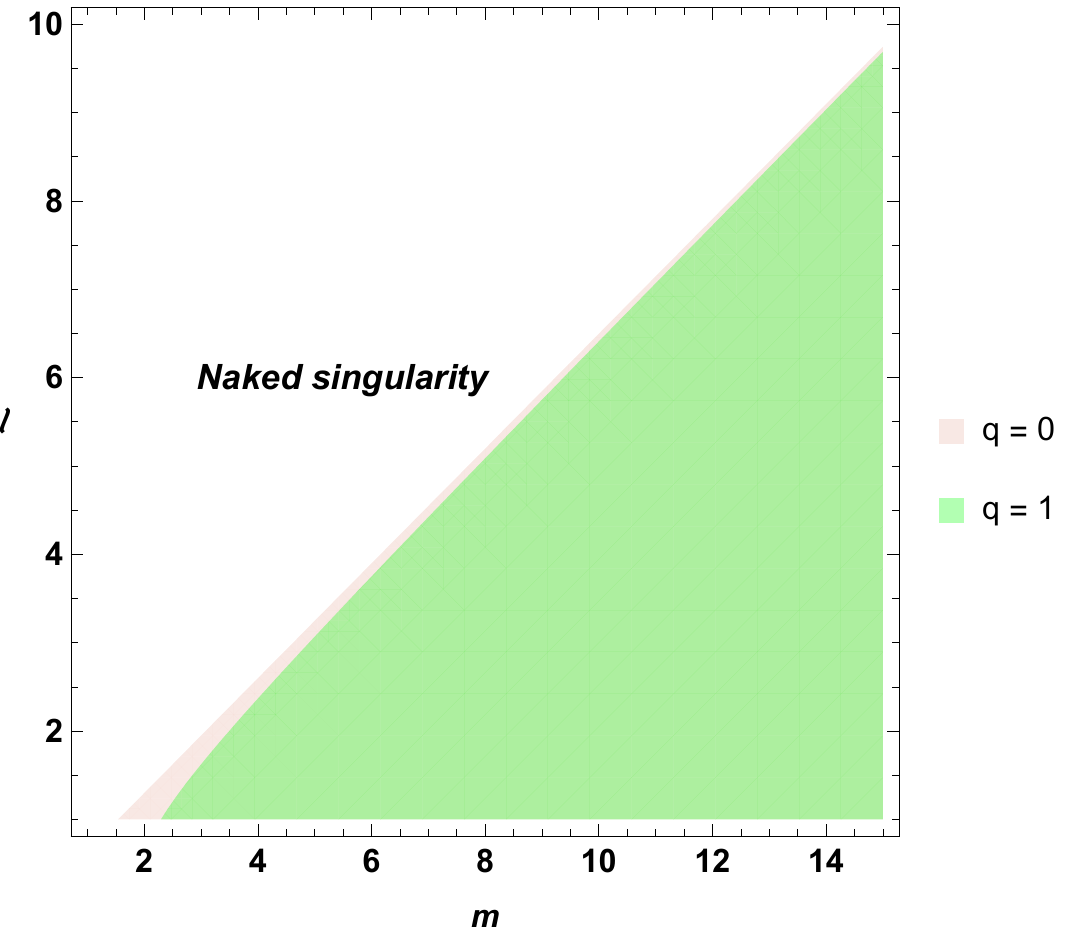} \\
		   \end{tabbing}
\caption{{\it \footnotesize {Left the shadow behaviors  of superentropic AdS black holes for different values of  $m$, and  $\ell$ by taking $q=0$, the right the region plot for the superentropic AdS black holes as function of parameters $l$, $m$ and $q$.The observer is positioned at $r_{ob}=100$ and $\theta_{ob}=\frac{\pi}{2}.$}}}
 \label{fa1}
\end{center}
\end{figure}
In Fig.(\ref{fa1}),  we illustrate the naked singularity behaviors for non-charged solutions by varying the mass parameter  in appropriate range of $\ell$.
It is observed  that the length of the arcs  depends   on  such parameter variations. In particular,    it is increasing with $m$ and $\ell$. It  has been observed that   the arcs  became closed  when the $m$ and  $\ell$ black hole  parameters increase.   This could understood from the mass variation.  In Fig.(\ref{fa1}), we plot the  horizon region   and  the naked singularity  for  such a black hole  as function of the parameters $m$ and $\ell$ for certain specific values of $q$.  In the absence of the charge,   the horizon  region  is relevant with respect to charged solutions. It has been remarked generic values of $q>1$, the difference is significant.

\vspace{4cm}
\section{Conclusion}
In this work,  we have investigated   shadow  behaviors of superentropic black holes in four dimensions. It has been shown that the mass constraint have generated two different solutions associated with superentropic black holes, being existing horizons and naked singularities. Applying  the  Hamilton-Jacobi  method,  we  have first found the    null  geodesic equations.  In  the celestial  coordinate  framework associated with fixed positions of observers,   we   have   engineered  the   shadow  geometries of existing horizons  by  varying the mass and  the cosmological scale  parameters.  Among others, the size shadow augments with the mass. However, it decreases with  the cosmological scale length.   Concretely, we    have obtained    ellipse shaped geometries contrary to  usual  black hole solutions.  For fixed values of $\ell$, the ellipse D-shape geometry appears for  particular mass values.   This  result could be  supported by the fact that such   black holes, with the mass constraint,  involve  a fast spinning motion.
Modifying the  ordinary   relations describing geometrical observables, we  have  examined      the size and  the distortion parameters of  such  elliptic  forms. To make a distinction between such two possible solutions, we have investigated the geometrical behavior of the shadow associated with the horizon-less configuration. In this way, the shadow  geometries have been given in terms of arcs depending on black holes physical parameters. The size of these arcs  increases with the mass and the AdS radius. It has been observed that these parameters act on such geometrical configurations in a similar way. \\
This work  comes up with certain questions. A natural question  concerns higher dimensional solutions which could be dealt   with in  alternative   theories  including stringy models. Such a  subject could open new roads to unveil significant physical data on rotating black hole physics.  Precisely, this could bring new  candidate geometries to model realistic black  holes in four   dimensions. It should be interesting to implement new physical aspects to study distinctions between horizon existing and horizon-less solutions.  We believe that these observations deeper reflections. These left for future works.

\section*{Acknowledgements}
The authors would like to thank A. El Balali, W. El Hadri, H. El Moumni, Y. Hassouni, M. B. Sedra, Y. Sekhmani,  M. Oulaid, E. Torrente   Lujano for discussions on related topics
and correspondence.  MB would like to thank Di Wu  for emailing  discussions  about the present work. This work is partially supported by the ICTP through
AF.

\end{document}